\documentclass[letterpaper, 10 pt, conference]{ieeeconf}  

\IEEEoverridecommandlockouts                              

\usepackage{amsmath} 
\usepackage{amssymb}  
\usepackage{algorithm}
\usepackage[noend]{algpseudocode}
\usepackage{color}
\usepackage{graphicx}
\usepackage{booktabs}
\usepackage{xcolor}

\title{\LARGE \bf State Constrained Stochastic Optimal Control Using LSTMs\thanks{This work is supported in part by the NSF under grant DMS-1907518.}}

\author{Bolun Dai$^{1}$, Prashanth Krishnamurthy$^{1}$, Andrew Papanicolaou$^{2}$, Farshad Khorrami$^{1}$
\thanks{$^{1}$Control/Robotics Research Laboratory, Electrical~\&~Computer Engineering Department, Tandon School of Engineering (Polytechnic Institute), New York University, Brooklyn, NY, 11201
{\tt\small bd1555@nyu.edu, prashanth.krishnamurthy@nyu.edu, khorrami@nyu.edu.}
}
\thanks{$^{2}$Department of Mathematics, College of Sciences, North Carolina State University, Raleigh, NC, 27695
{\tt\small apapani@ncsu.edu}.
}
}

\begin{document}
\maketitle
\thispagestyle{empty}
\pagestyle{empty}
\begin{abstract}
In this paper, we propose a new methodology for state constrained stochastic optimal control (SOC) problems. The solution is based on past work in solving SOC problems using forward-backward stochastic differential equations (FBSDE). Our approach in solving the FBSDE utilizes a deep neural network (DNN), specifically Long Short-Term Memory (LSTM) networks. LSTMs are chosen to solve the FBSDE to address the curse of dimensionality, non-linearities, and long time horizons. In addition, the state constraints are incorporated using a hard penalty function, resulting in a controller that respects the constraint boundaries. Numerical instability that would be introduced by the penalty function is dealt with through an adaptive update scheme. The control design methodology is applicable to a large class of control problems. The performance and scalability of our proposed algorithm are demonstrated by numerical simulations. 
\end{abstract}
\section{Introduction}
Optimal control has wide applications in robotic control~\cite{DBLP:conf/iros/HerdtPW10}, navigation~\cite{DBLP:conf/iros/CassandraKK96}, to name a few. With the availability of higher computational power and powerful optimization software, such as SNOPT~\cite{GilMS05}, optimization algorithms have been applied to increasingly complex control problems. Optimal control casts the control problem in terms of cost functions that are addressed through numerical optimization techniques. In reality, systems also contain many uncontrolled inputs, such as measurement noise and external forces with unknown distributions, which if not taken into consideration in design, will lead to performance degradation. To address these uncontrolled inputs, a set of optimal control problems, namely stochastic optimal control, have been considered, which takes disturbances or measurement noise into consideration more explicitly.

The solution to a SOC problem involves solving a nonlinear partial differential equation (PDE), known as the Hamiliton-Jacobi-Bellman (HJB) equation. One approach is to approximate the state dynamics by linear~\cite{1469949} or quadratic~\cite{5530971} systems. However, this requires fine-grained time discretization or specialized linearization techniques. When the system is of high dimension, solving the HJB equation becomes difficult due to the curse of dimensionality. An alternative to linearization is sampling-based methods, such as Markov-Chain Monte Carlo (MCMC) approximation~\cite{doi:10.1177/0278364915616866} to the HJB equation and forward-backward stochastic differential equations (FBSDE)~\cite{EXARCHOS2018159}. However, numerical methods for MCMC-based methods are difficult to scale due to their reliance on predefined grids for value function backward propagation~\cite{doi:10.1177/0278364915616866}. They also suffer from compounding errors in least square approximations for FBSDE-based methods~\cite{EXARCHOS2018159}. Work has also been done in state constrained SOC problems,~\cite{doi:10.1137/15M1023737} translated the problem into a state-constrained target problem and used its backward reachable sets to describe the value function. However, the efficacy of~\cite{doi:10.1137/15M1023737} was not demonstrated on a physical system.

Deep learning has been successfully applied to several application domains in recent years. One application of deep learning is to serve as an alternative to numerical methods in solving high-dimensional and nonlinear PDEs. In~\cite{Han8505}, DNNs are used to solve PDEs, such as Black-Scholes, HJB, and Allen-Cahn equations. However, in~\cite{Han8505}, only problems with available analytic solutions and uncontrolled state dynamics are tested. Following this line of work, DNNs have been utilized to solve FBSDEs for SOC problems. By using DNNs, we can learn the initial condition and then propagate forward the solution to the backward stochastic differential equation (BSDE). Additionally, solving the FBSDE using neural network (NN) based methods resolves the aforementioned compounding-error issue. The basic formulation of using LSTMs to solve FBSDEs was shown in~\cite{DBLP:conf/rss/WangPT19}, along with adding control constraints. However, the paper did not consider state constraints, which occur in most physical control problems. Another branch of deep learning, namely deep reinforcement learning (DRL), has also been deployed to solved optimal control problems. One significant difference between deep FBSDE based methods and DRL is the control in deep FBSDE based methods are given as functions of the value function whereas in DRL, the value function is either used as a baseline to reduce variance~\cite{DBLP:journals/corr/SchulmanWDRK17} or a state-action value function is used~\cite{DBLP:journals/corr/MnihKSGAWR13} to obtain control actions.

In this paper, we propose a SOC setting for nonlinear systems that incorporates state constraints. In Section II, the state constrained SOC formulation is given. In Section III, the deep FBSDE algorithm is presented along with modifications to include control saturation. In Section IV, an algorithmic solution is provided for adding state constraints. This results in a log-barrier-like method. In Section V, the NN architecture and overall algorithm are presented. In Section V, an adaptive update scheme to the state constraint penalty function is shown, which enhances stability during NN training. In Section VI, we show the efficacy of our approach on a cart-pole system under two different state constraints, namely cart movement and system energy. Finally, the paper is concluded and potential future research directions are presented.
\section{Problem Formulation}
In this section, we outline the optimal control problem under state constraints. A system with dynamics that involves stochastic processes can be described using a stochastic differential equation (SDE) as follows
\begin{equation}
	\label{eq:SDE2}
	dx(t) = f(x(t),t)dt + G(x(t),t)u(t)dt + \Sigma(x(t),t)dw(t)
\end{equation}
with initial condition $x(0) = x_0$ and $w(t)\in\mathbb{R}^\nu$ being a standard Brownian motion. The states are denoted by $x\in\mathbb{R}^n$, time by $0<t<T<\infty$, and the control input by $u\in\mathbb{R}^m$. In \eqref{eq:SDE2}, $f: \mathbb{R}^n\times[0, T]\rightarrow\mathbb{R}^n$ represents the drift, $G: \mathbb{R}^n\times[0, T]\rightarrow\mathbb{R}^{n\times m}$ represents the control influence, and $\Sigma: \mathbb{R}^n\times[0, T]\rightarrow\mathbb{R}^{n\times\nu}$ represents the diffusion (influence of the Brownian motion on the state evolution). The state constrained SOC problem is to find a controller $u(t)$ that minimizes an objective function $J^u(x,t)\in\mathbb{R}^+$ under a set of state constraints. The objective function is defined as
\begin{align}
    J^u(x,t) =&\ \mathbb{E}\Big[g(x(T)) + \int_t^T\Big(q(x(s))\nonumber\\
              &+\frac{1}{2}u(s)^TRu(s)\Big)ds\Big|x(t) = x\Big]
    \label{eq:objective_optimization}
\end{align}
where $g:\mathbb{R}^n\rightarrow\mathbb{R}^+$ is the terminal state cost, $q:\mathbb{R}^n\rightarrow\mathbb{R}^+$ is the instantaneous state cost, and the control cost matrix is $R\in\mathbb{R}^{m\times m}$, which is positive definite. We consider state constraints in the following form
\begin{equation}
    c_{\min} \leq c_s(x) \leq c_{\max}
\end{equation}
where $c_s(x)\in\mathbb{R}^r$ is a vector of functions of the state, and $c_{\min}\in\mathbb{R}^r$ and $c_{\max}\in\mathbb{R}^r$ represent the lower and upper bounds of $c_s(x)$ component-wise, respectively. An example of this kind of constraint is a box constraint on states, i.e., wherein $c_s(x) = x$. Additionally, control saturation (with $U_{\max}\in\mathbb{R}^m$) is introduced, which has the form of
\begin{equation}
    u\in\mathcal{U} = \{u\ |\ |u_i|\leq U_{i, \max}\}
    \label{eq:control_saturation}
\end{equation}
where $u_i$ and $U_{i, \max}$ are the $i^{th}$ elements of $u$ and $U_{\max}$, respectively. 
The value function $V(x,t)\in\mathbb{R}^+$ is defined as
\begin{equation}
    V(x,t) := \inf J^{u}(x,t)
\end{equation} 
where the $\inf$ is computed over all control signals $u(.)$ over the time interval $(t, T]$ satisfying the constraint \eqref{eq:control_saturation}.

\section{Deep FBSDE Formulation}
To solve the state constrained SOC problem, a FBSDE formulation is used. In this section, we first consider the unconstrained SOC problem. Later in this section, we introduce the deep FBSDE formulation that is modified to handle control saturation. Modifications required to handle state constrained SOC problems will be introduced in Section~\ref{sec:state_constraint}. Starting from~\eqref{eq:objective_optimization}, we can use Dynkin's formula~\cite{dynkin1965} along with Bellman's principle~\cite{10.5555/862270} to arrive at the HJB equation
\begin{subequations}
\begin{align}
    & V_t(x,t) + \mathcal{L}V(x,t) + h(x,V_x,t)\\
    & V(x,T) = g(x)
\end{align}
\label{eq:HJB}
\end{subequations}
\noindent 
\hspace*{-0.25cm} where $V_t$ is the partial derivative of $V$ with respect to $t$. The generator function $\mathcal{L}V(x,t)$ is defined as
\begin{equation}
    \mathcal{L}V = \frac{1}{2}\hbox{trace}(\Sigma\Sigma^TV_{xx}) + f^TV_x
\end{equation}
where $V_x$ and $V_{xx}$ denote the first and second order partial derivatives of $V$ w.r.t. $x$, respectively. The Hamiltonian is
\begin{equation}
    h(x,V_x,t) \!=\! \inf_{u\in\mathcal U}\left(q(x)+(G(x,t)u)^TV_x(x,t)+\frac{1}{2}u^TRu\right).
\end{equation}
Using first-order conditions to solve for the optimal control in the Hamiltonian yields
\begin{equation}
	\label{eq:feedbackForm}
	u^*(x,t) = -R^{-1}G^T(x,t)V_x(x,t).
\end{equation}
Define $G(x(t), t) = \Sigma(x(t), t)\Gamma(x(t), t)$, where $\Gamma:  \mathbb{R}^n\times[0, T]\rightarrow\mathbb{R}^{\nu\times m}$. 
This form of $G$ captures the characteristic that the range of $G$ belongs to the range of $\Sigma$, which excludes the case of a channel containing control input without noise. This representation aligns with the fact that no noiseless control signal exists in reality.

Additional to this basic formulation, the control inputs are saturated as shown in \eqref{eq:control_saturation}. The saturation is introduced using the $\mathrm{sig}$ function as in~\cite{DBLP:conf/rss/WangPT19}, which is defined as
\begin{equation}
    \mathrm{sig}(v) = \frac{2}{1 + e^{-v}} - 1.
\end{equation}
The optimal control action is then calculated as
\begin{equation}
    {u}^*(x, t) = U_{\max}*\mathrm{sig}(-R^{-1}G^T(t, x)V_x)
    \label{eq:control_constraint}
\end{equation}
with the control saturated between $[-U_{\max}, U_{\max}]$. In \eqref{eq:control_constraint}, ``$*$" represents element-wise multiplication. For the control saturated system, a new value function is introduced
\begin{equation}
    \mathbb E\left[g(x(T)) + \int_t^T \! \! \left(q(x(s)) + \sum_{i = 1}^{m}{S_i(u_i(s))}\right)ds\Big|x(t) = x\right].
\end{equation}
The new control cost $S_i(u_i)$ is defined as
\begin{equation}
    S_i(u_i) = c_i\int_{0}^{u_i}{\mathrm{sig}^{-1}\Big(\frac{v}{U_{i, \max}}\Big)dv}
    \label{eq:control_constrain_cost}
\end{equation}
with constant weights $c_i$, dummy variable for integration $v$. For this control saturated system, the Hamiltonian will become
\begin{equation}
    h(x, V_x, t, u^*) = q(x) + V_x^TG(x, t)u^*(x, t) + \sum_{i = 1}^{m}{S_i(u_i^*)}.
    \label{eq:cc_hamiltonian}
\end{equation}

Following the formulation in~\cite{DBLP:conf/rss/WangPT19}, we can solve for ~\eqref{eq:control_constraint} using a FBSDE, which has the form of
\begingroup
\allowdisplaybreaks
\begin{subequations}
\label{eq:FBSDE_dnn}
\begin{align}
dy(t) =&\ \Big(-h\big(x(t), \Sigma^T(x(t),t)V_x(x(t),t; \theta), t\big)\nonumber\\
        &+ V_x^T\big(x(t), t; \theta\big)G\big(x(t),t\big)u\big(x(t),t\big)\Big)dt\nonumber\\ 
        &+ V_x^T\big(x(t),t; \theta\big)\Sigma\big(x(t),t\big)dw(t)\\
dx(t) =&\ \Big(f\big(x(t),t\big) +G\big(x(t),t\big)u(x(t),t\big)\Big)dt\nonumber\\
       &+ \Sigma\big(x(t),t\big)dw(t)\\
u(t) =&\ U_{\max} * \mathrm{sig}(-R^{-1}G^T\big(x(t), t\big)V_x(x(t),t; \theta)\\
y(0) =&\ V\big(\phi\big)\\
dy(0) =&\ V_x\big(\phi\big)\\
x(0) =&\ x_0.
\end{align}
\end{subequations}
\endgroup
The deep FBSDE method~\cite{DBLP:conf/rss/WangPT19} introduces a NN with parameters $\theta$ to estimate $V_{x}(x(t_n), t_n)$ at each time step, which is denoted as $V_{x}(x(t_n), t_n; \theta)$. Using the $V_{x}(x(t_n), t_n)$ estimation and following~\eqref{eq:control_constraint}, we can obtain the control action $u_n^*$. The initial value function estimation $V(x(t_0), t_0)$ and its partial derivative $V_{x}(x(t_0), t_0)$ are also learned via trainable weights $\phi$; with some abuse of notation, we denote these learned models as $V(\phi)$ and $V_x(\phi)$. This enables the forward propagation of the backward part in the FBSDE. The original continuous finite time horizon problem is discretized into $N$ time steps ($T = N\Delta{t}$). This gives us $\Delta y(t_n)$ and $\Delta x(t_n)$. Using a numerical integration scheme as in~\cite{Han8505}, we can find the value function estimation at the last time step $V(x(t_N), t_N)$ along with the terminal state $x(T)$. The loss function of the DNN is chosen such that $\|V(x(t_N), t_N) - g(x(T))\|_2$ is minimized. Note that $V(x(t_N), t_N)$ is equivalent to $y(T)$.

\section{State Constraint Formulation}
\label{sec:state_constraint}
\begin{figure}[t]
    \centering
    \includegraphics[width=0.49\textwidth]{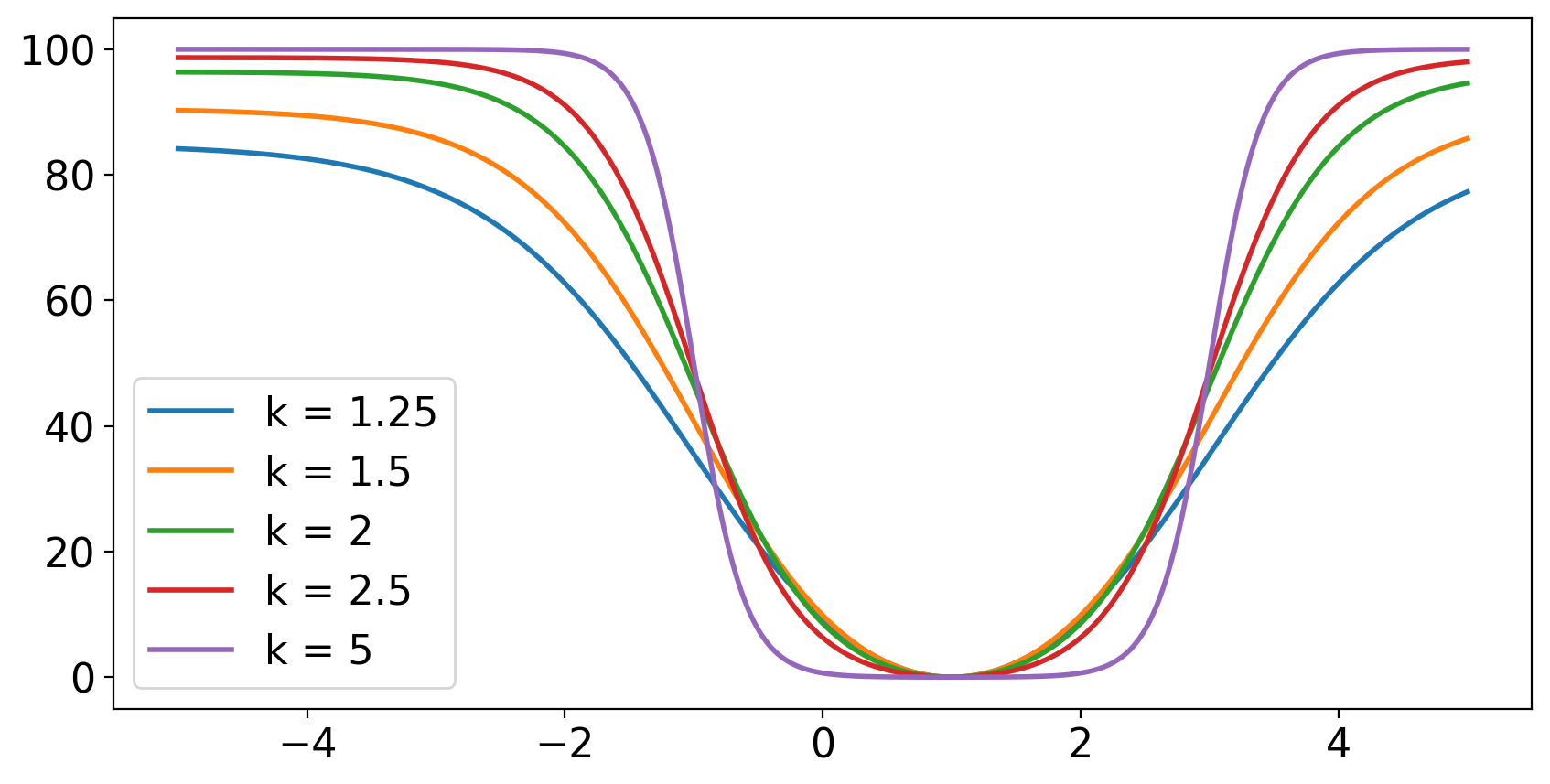}
    \caption{$p(x)$ in~\eqref{eq:soft_constrain} under different values of $k$, with $\mu = 1$, $L = 100$, $c_s(x) = x$, $c_{\min} = -1$, $c_{\max} = 3$ and $x$ being a scalar. It can be seen that as $k$ increases, the boundaries gets increasingly steep, and the values of $p(x)$ inside the constrained region $(-1, 3)$ go to zero.}
    \label{fig:soft_constrain}
\end{figure}

In this section, we present our formulation to incorporate state constraints on top of the control saturated FBSDE formulation. At each time instant, a NN is used to estimate $V_x$. Then, $V_x$ is used to calculate the optimal control that minimizes the cost-to-go. Additionally, the optimal controller of the state constrained SOC problem needs to satisfy the state constraints. We can modify the cost function, such that the cost-to-go is minimized only when the state constraints are respected. This can be achieved by adding a ``soft" constraint in the form of a penalty, which increases the cost when the state is outside the constraint boundary. This approach is preferred over using a sigmoid-like function to force the state dynamics into a predefined range. Using the $\mathrm{sig}$ function, the dynamics would be altered using a “fake” saturation which is not present in the real dynamics — hence, training on such an altered dynamic model would not yield a controller that functions properly under the real dynamics. When adding the penalty, states inside the constraint boundary should not be penalized, but for states outside the constraint boundary, a large penalty should be given. Therefore, the penalty function should be close to zero inside the constraint boundary and a large number outside. The function
\begin{equation}
\begin{aligned}
    p(x) =&\ \frac{L}{1 + e^{-k(c_s(x) - c_{\max})}} - \frac{L}{1 + e^{-k(c_s(x) - c_{\min})}}\\
          &+ L - \frac{2L}{1 + e^{-k(\mu - c_{\max})}}
\end{aligned}
\label{eq:soft_constrain}
\end{equation}
satisfies the aforementioned requirements. $L\in\mathbb{R}^+$ is a scalar determining the maximum value of the penalty, $k\in\mathbb{R}^+$ determines the steepness of the boundary (larger $k$ leads to steeper boundaries), as shown in Figure~\ref{fig:soft_constrain}, and $\mu = 0.5(c_{\max} + c_{\min})\in\mathbb{R}^r$ represents the mid-point of the constraint region. The proposed penalty function consists of two parts: the first part is the two logistics functions, which gives the ``U"-like shape; the second part moves the minimum value of the penalty function to zero.

Using the penalty function $p(x)$, the new state cost at each time instant becomes
\begin{equation}
    c(x) = q(x) + p(x).
    \label{eq:state_cost}
\end{equation}
Note, when choosing parameters for $p(x)$, ideally we would pick larger $k$ and $L$ values to ensure a steeper boundary. After applying a penalty function, the Hamiltonian becomes
\begin{equation}
    h(x, V_x, t, u^*) = c(x) + V_x^TG(x, t)u^*(x, t) + \sum_{i = 1}^{m}{S_i(u_i^*)}
    \label{eq:sc_hamiltonian}
\end{equation}
\noindent where $u^*$ is defined in~\eqref{eq:control_constraint}. Thus, state constraints can be applied by simply swapping the Hamiltonian defined in~\eqref{eq:cc_hamiltonian} to the form in~\eqref{eq:sc_hamiltonian}.
\begin{algorithm}[t]
\caption{Soft state constraint update}
\begin{algorithmic}[1]
    \State {\bf Given}:
    \State $k$: Boundary steepness, $\delta$: Change of boundary steepness, $\beta$: Update threshold, $\gamma$: Threshold change ratio, $l$: Iteration number, $\Delta$: Boundary steepness change acceleration, $\eta$: Update interval, $\eta'$: Max interval, $\Delta_\delta$: Threshold change acceleration;
    \If{state trajectory not inside constraint boundary}
        \If{$l\ \mathbf{mod}\ \eta = 0$}
            \State $\bar{c} =  \frac{1}{N}\sum_{l=1}^{\eta}{c(x_l)}$;
            \State $\sigma_c^2 = \frac{1}{N}\sum_{l=1}^{\eta}{(c_l - \bar{c})^2}$;
            \If{$\sigma_c < \beta$ or $l\ \mathbf{mod}\ \eta' = 0$}
                \State $k, \delta, \beta, \gamma = k + \delta, \delta - \Delta_\delta, \gamma\beta, \gamma + \Delta$
            \EndIf
        \EndIf
        \If{$\delta < 0$} 
            \State $\delta = 0$;
        \EndIf
        \If{$\gamma > 1$}
            \State $\gamma = 1$;
        \EndIf
    \Else
        \State End Algorithm
    \EndIf
\end{algorithmic}
\label{alg:k_update}
\end{algorithm}

\section{State Constrained Deep FBSDE Controller}
In this section, we present the state constrained deep FBSDE algorithm, an overview of the corresponding NN architecture, and an adaptive update scheme to the state constraints that greatly enhances training stability.
\subsection{Algorithmic Design}
The time horizon $T$ is discretized into $N$ time steps. At each time step $n$, we use a NN parameterized by $\theta$. The input to the NN at each time step is the current state $x_n^m$ (and the hidden state from the previous time step $H_{n-1}^m$ when using LSTMs), where $m$ represents the $m$-th sample in the mini-batch. The output of the NN at each time step is $V_x^m(x(t_n), t_n))$ , which is then used to calculate the control action using~\eqref{eq:feedbackForm}. Following~\eqref{eq:FBSDE_dnn}, we can obtain the next value function estimation and the next state using Euler's integration. After the last time step, the loss is computed as
\begin{equation}
    L = \frac{1}{M}\sum_{m=1}^{M}\|g(x_N^m) - y_N^m\|_2^2 + \lambda\|\theta\|_2^2
\end{equation}
where $\lambda$ determines the weight of the regularization term. The initial value function estimate $V(x(t_0), t_0)$ and its partial derivative $V_{x}(x(t_0), t_0)$ are parameterized by $\phi_V$ (if using LSTMs, initial hidden states $H_0$ are parameterized by $\phi_H$). The initial state $\xi$ is fixed. We use a batch size of $M$ during training. The NN is trained for a total $L$ iterations using the Adam optimizer under a varying learning rate. All of the weights are initialized using the Xavier normal initializer.

\subsection{Neural Network Architecture}
\begin{figure*}[t]
    \centering
    \includegraphics[width=\textwidth]{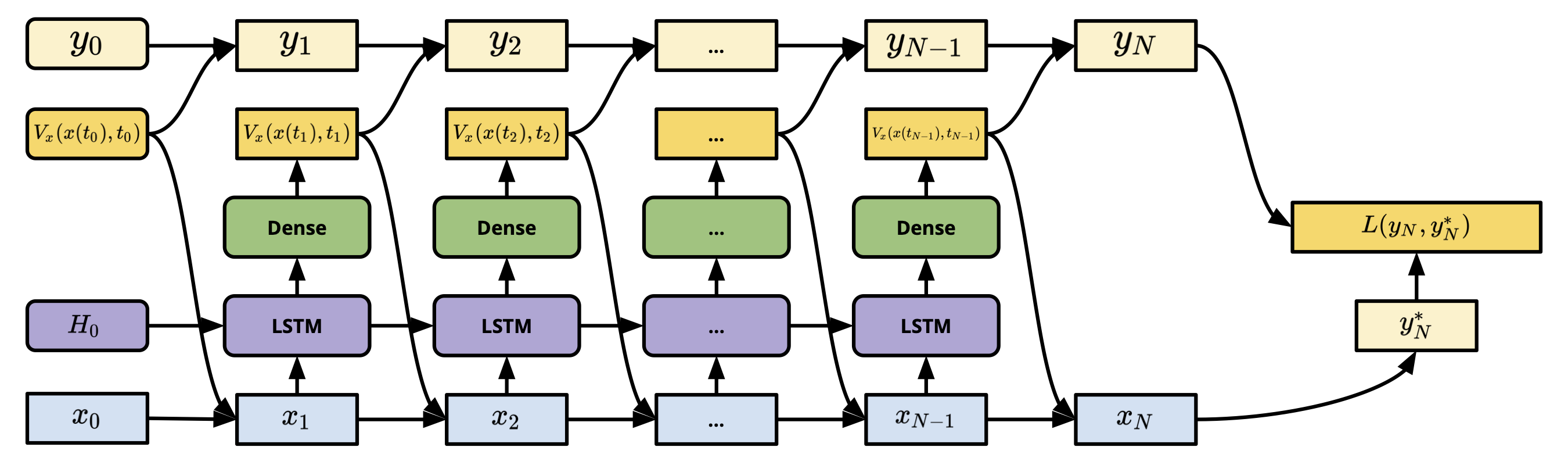}
    \caption{\textbf{Neural Network Architecture}: the architecture here is for $N$ timesteps, the connection between $V_{x}(x(t_n), t_n)$ and $x_{n+1}$ denotes applying control action, and the weights for each timestep are shared. Curved edge boxes represent trainable parameters, sharp edges represent intermediate values.}
    \label{fig:nn_arch}
\end{figure*}

\begin{figure*}[t!]
    \centering
    \includegraphics[width=\textwidth]{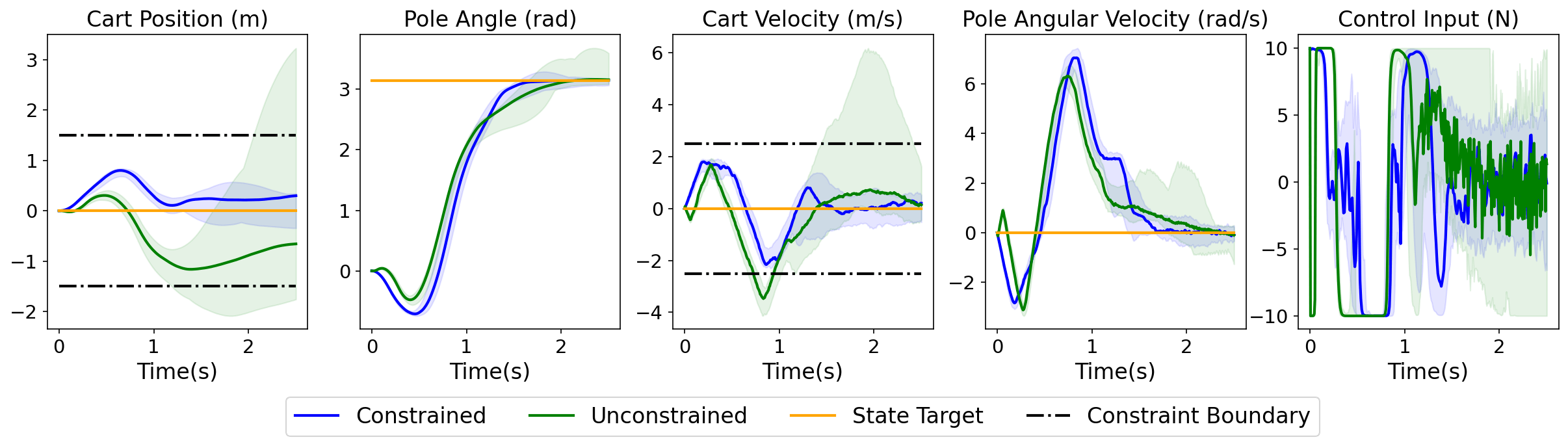}
    \caption{\textbf{Cart-pole State Dynamics}: performance comparison of the controller when the state is constrained (blue) and not constrained (green). The darker blue and green trajectories are sampled trajectories among the 256 trials. The light blue and green regions show the bounds of the maximum and minimum values of all 256 trajectories at each time step. The constraint boundaries are shown in black dashed lines. The orange line shows the target value for each of the states. All the state trajectories of the constrained system satisfy the constraint boundary. Without applying state constraints, the state trajectories (light green region) violate the constraint boundaries.}
    \label{fig:final_state}
\end{figure*}

\begin{figure*}[t!]
    \centering
    \includegraphics[width=\textwidth]{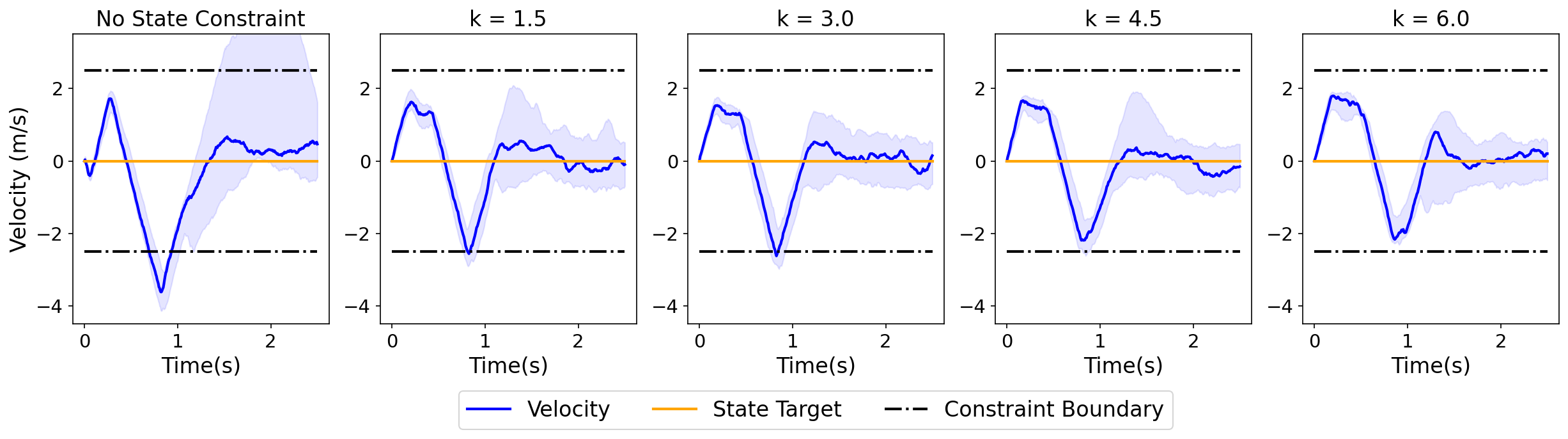}
    \vspace*{-0.4cm}
    \caption{\textbf{Effectiveness of Adaptive Update Scheme}: the color scheme for the velocity follows the constrained curve in Figure~\ref{fig:final_state}. The state target and constraint boundary also follows the color scheme in Figure~\ref{fig:final_state}. The leftmost figure shows that under no state constraints the cart velocity violates the constraints. After increasing $k$, the velocity trajectory gradually retreats inside of the boundaries and eventually, it entirely satisfies the constraints.}
    \label{fig:k_change}
    \vspace*{-0.1in}
\end{figure*}

Both dense-layer based and LSTM based NN architectures have been used in~\cite{DBLP:conf/rss/WangPT19}. The benefits of using a LSTM based architecture are two-fold. Firstly, the weights are shared among time steps, compared to a dense-layer based architecture resulting in reduced number of weights. Secondly, for dense-layer based methods, weights at each time step are step specific, making it difficult to scale to long horizons. For LSTM based architectures, we can train for a shorter time horizon than during testing, given the state distribution for longer time horizons are the same. The NN architecture illustrated in Figure~\ref{fig:nn_arch} is inspired by~\cite{DBLP:conf/rss/WangPT19}; thus, only a brief introduction is provided. A LSTM-based architecture is used to tackle the vanishing gradient problem in long time horizons. Additional to the basic LSTM model, we have a forward part which represents the flow of state dynamics and a backward part representing the flow of the value function.

\subsection{Adaptive Update Scheme}
In the initial stage of training, parts of many trajectories lie outside of the constraint boundary. If a steep boundary is in use, it will result in a very large gradient and cause numerical instabilities in training even if the gradient is clipped. If the steepness of the boundary is fixed at a small value, the penalty of violating the constraint is not significant enough for the learning algorithm to achieve the required state constraint. Therefore, an adaptive update scheme is needed to gradually increase the boundary steepness $k$ in order to avoid exploding gradients while shaping the controller behavior.

For a given $k$, we train the DNN until the performance of the controller cannot be further improved, then we update $k$. To determine the performance of the controller, we can use the square root of state cost variance $\sigma_c\in\mathbb{R}$ over a given amount of iterations. The state cost is defined in~\eqref{eq:state_cost}; its square root variance decreases as the learning algorithm converges to a solution. Define the threshold of determining convergence as $\beta\in\mathbb{R}^+$. Once $\sigma_c$ becomes smaller than $\beta$, we update $k$. We check for the condition $\sigma_c < \beta$ every $\eta$ iterations; if the condition is not satisfied after $\eta'$ iterations, $k$ will also be updated (may be stuck in a local minimum). 

During the training process, the variance decreases. Thus, $\beta$ should also be decreased after each update. To make the decrease in $\beta$ smoother, we let the ``acceleration," $\Delta\in\mathbb{R}^+$, become negative. Similarly, the acceleration of the increase of $k$ is made negative, which is denoted by $\Delta_\delta\in\mathbb{R}^+$, leading to more fine-grained changes in $k$ at later stages of training. Algorithm~\ref{alg:k_update} shows the update scheme for $k$ and the overall algorithm is given in Algorithm~\ref{alg:fbsde}.
\begin{algorithm}[t!]
\caption{State Constrained Deep FBSDE Controller}
\begin{algorithmic}[1]
    \State {\bf Given}:
    \State $x_0 = \xi$, $f(t, x)$, $G(t, x)$, $\Sigma(t, x)$: Initial state and state dynamics;
    \State $g(x)$, $\nabla_xg(x)$, $q(x)$, $R$: Cost function parameters;
    \State $N$: Task horizon, $K$: Number of iterations, $M$: Batch size, $\Delta{t}$: Time interval, $\lambda$: Weight decay parameter;
    \State $\Delta$, $\eta$, $\eta'$, $\Delta_\delta$: state constraint parameters (refer to Algorithm~\ref{alg:k_update});
    \State {\bf Initialization}:
    \State $\theta$: Weights and biases of dense and LSTM layers;
    \State $\phi$: Weight of the initial value function and hidden unit;
    \State $k$, $\delta$, $\beta$, $\gamma$ (refer to Algorithm~\ref{alg:k_update});
    \State $\{x_0^m\}_{m=1}^M = \{\xi\}_{m=1}^M$;
    \State $\{y_0^m\}_{m=1}^M = \{V(\phi)\}_{m=1}^M$: Value function at $t = 0$;
    \State $\{V_{x}^m(x(t_0), t_0)\}_{m=1}^M = \{V_{x}(\phi)\}_{m=1}^M$: Gradient of value function at $t = 0$;
    \State $\{H_{0}^m\}_{m=1}^M = \{H_{0}(\phi)\}_{m=1}^M$: Hidden units at $t = 0$;
    \State $t = 0$;
    \For {$l = 1$ to $L$}
        \For{$m = 1$ to $M$}
            \For{$n = 0$ to $N-1$}
                \State $u_n^m = -R^{-1}G^T(t, x_n^m)V_{x}^m(x(t_n), t_n)$
                \State ${u_n^m}^* = U_{\max}*\mathrm{sig}(u_n^m)$;
                \State Sample Brownian noise: $\Delta w_n^m\sim\mathcal{N}(0, \Delta{t})$;
                \State Update value function: 
                \State $y_{n+1}^m = y_n^m - h(t, x_n^m, y_n^m, V_{x}^m(x(t_n), t_n))\Delta{t} + (V_{x}^m(x(t_n), t_n))^T(G(t, x_n^m){u_n^m}^*\Delta{t} + \Sigma(t, x_n^m)\Delta w_n^m)$;
                \State Update system state:
                \State $x_{n+1}^m = x_n^m + f(t, x_n^m)\Delta{t} + G(t, x_n^m){u_n^m}^*\Delta{t} + \Sigma(t, x_n^m)\Delta w_n^m$;
                \State Predict gradient of value function:
                \State $V_{x}^m(x(t_{n+1}), t_{n+1}) = f_\mathrm{LSTM}(x_{n+1}^m, \theta^k)$;
            \EndFor
            \State Compute target terminal cost: $(y_N^m)^* = g(x_N^m)$;
        \EndFor
        \State Compute mini-batch loss:
        \State $\displaystyle L = \frac{1}{M}\sum_{i=1}^{M}\|(y_N^m)^* - y_N^m\|_2^2 + \lambda\|\theta^k\|_2^2$;
        \State Update $\theta$ and $\phi$ via backpropagation;
        \State $t = t + \Delta{t}$;
        \State Run Algorithm~\ref{alg:k_update};
    \EndFor
\end{algorithmic}
\label{alg:fbsde}
\end{algorithm}
\section{Experiments}
In this section, we show the efficacy of our control algorithm on the cart-pole for a swing-up task under two different state constraints: cart movement and system energy. For both systems, the trained model is evaluated over 256 trials. All experiments are carried out using TensorFlow.

\subsection{Cart-pole Swing-Up Task I}
\label{sec:cartpole1}
The task for the cart-pole system is to swing the pole up from downward position and stabilize it at the upright position. The cart-pole dynamics are given as
\begin{align}
    (M+m)\Ddot{x} + mL\sin\theta\Dot{\theta}^2 - mL\cos\theta\Ddot{\theta} &= u\\
    mL^2\Ddot{\theta} - mL\cos\theta\Ddot{x} - mgL\sin\theta &= 0
\end{align}
where $x$ is the cart position, $\theta$ is the pendulum angle with respect to the downward position, $M = 1.0$kg is the cart mass, $m = 0.01$kg is the pole mass (point mass at the tip), and  the pole length $L$ is $0.5$m. The initial state is at origin and the target state is $[0, \pi, 0, 0]^T$, with the states being $[x\ \theta\ \Dot{x}\ \Dot{\theta}]^T$. The control input $u$ is saturated at $\pm10$N. The cart position $x$ and velocity $\dot{x}$ are constrained at $\pm1.5$m and $\pm2.5$m/s, respectively. The time horizon is chosen to be $2.5$ sec and the time step is $\Delta{t} = 1/110$ sec. Disturbances are applied to the linear and angular velocities with a discount factor of 0.25. We use the cost function
\begin{equation}
    \sum_{i=1}^{N}\Big[\frac{1}{2}\mathbf{X}_i^TQ\mathbf{X}_i + S_i(u_i) + p(\mathbf{X}_i)\Big]
    \label{eq:exp_cost}
\end{equation}
where $\mathbf{X}_i$ is the difference between the current state and the target state. $Q$ is the cost weight matrix for the state, and $S_i$ is defined in~\eqref{eq:control_constrain_cost}. $p(\cdot)$ is the penalty function that incorporates the state constraint. The state dynamics under the control generated by the trained model are shown in Figure~\ref{fig:final_state}. For the constrained stochastic optimal controller, the state lies within the constraint boundary for all time. Figure~\ref{fig:k_change} shows that without the presence of a state constraint, the cart position and velocity do violate constraints significantly. After applying the state constraint, the effect can be immediately seen on the state trajectories. With $k = 1.5$, the portion of the trajectory that lies outside of the constraint boundaries is greatly reduced. After gradually increasing $k$ to $6.0$ following Algorithm~\ref{alg:k_update}, the entire trajectory of the cart velocity lies within the constraint boundaries. Also, we tested directly training on $k = 6.0$ without such gradual increase, the algorithm becomes numerically unstable after one forward pass due to large gradients. This demonstrates that our method provides a stable training scheme.

\subsection{Cart-pole Swing-Up Task II}
\begin{figure}[t!]
    \centering
    \includegraphics[width=0.5\textwidth]{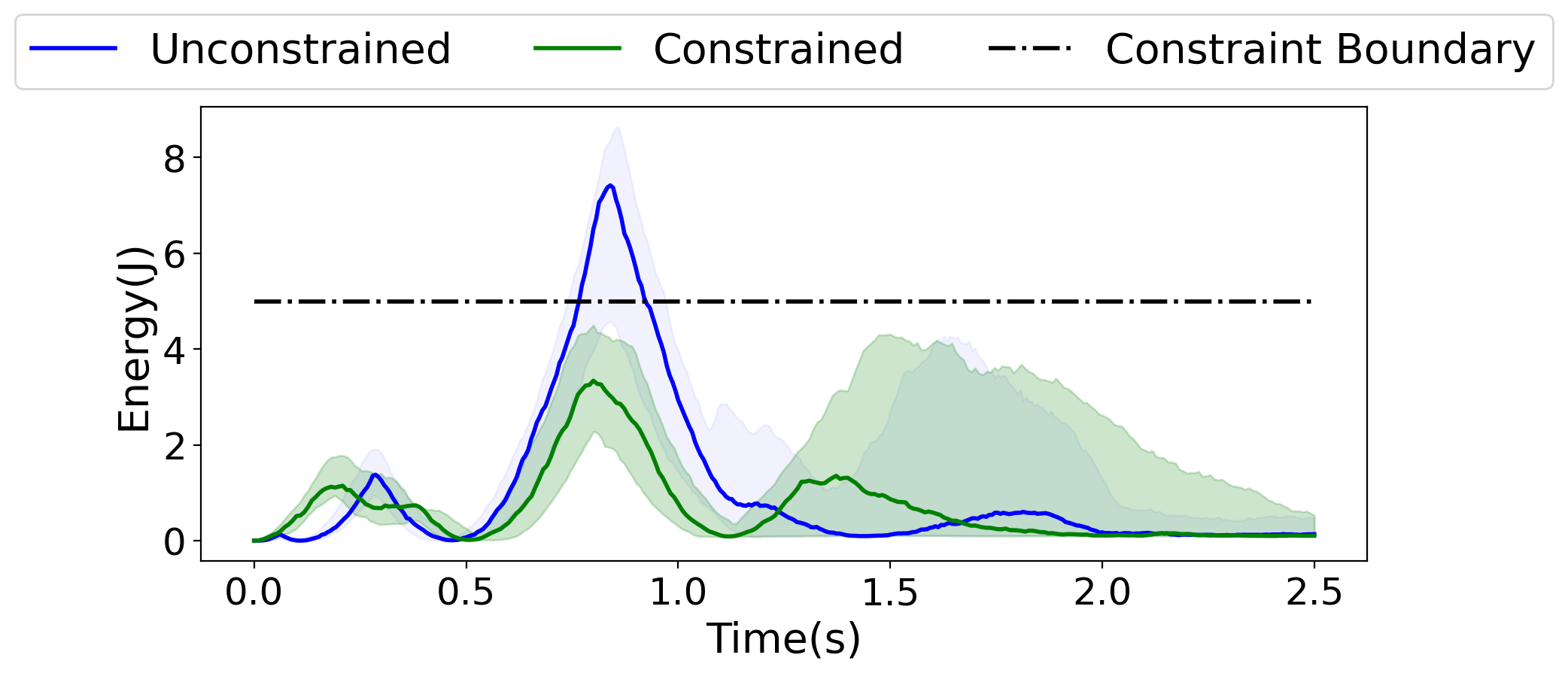}
    \caption{\textbf{Energy Constrained Cart-pole Controller}: the color scheme follows the one from Figure~\ref{fig:final_state}. It can be shown that under the constrained controller the total energy is able to be maintained under $5J$ which is not achieved under the unconstrained controller.}
    \label{fig:energy_comparison}
    \vspace*{-0.1in}
\end{figure}

Using the same model as before, a state constraint is applied to limit the total energy of the system (sum of potential and kinetic energy), which is a nonlinear function of the states. For the cart-pole system, the total energy
\begin{equation}
    E = \frac{1}{2}M\dot{x}^2 + mgL(1 - \cos\theta) + \frac{1}{2}mL^2\dot{\theta}^2
\end{equation}
is constrained at $\pm5J$. The cost is similar to~\eqref{eq:exp_cost}, with the only difference being changing $p(\cdot)$ to incorporate the system energy constraint. The total energy under the unconstrained and constrained controllers are shown in Figure~\ref{fig:energy_comparison}. 

 \vspace*{0.05cm}
\section{Conclusion}
In this paper, SOC problems with state constraints are formulated as a FBSDE and solved utilizing an LSTM-based DNN to alleviate the curse of dimensionality and numerical integration issues. An adaptive update scheme is used to apply state constraints, which greatly enhanced the stability during training of the LSTM network. The efficacy of our approach is demonstrated on a cart-pole system under two different state constraints: 1) cart position and velocity, and 2) system energy. Potential future directions include application to systems with partial state measurements and robust control under various uncertainties.

\vspace*{-0.05in}
\bibliographystyle{IEEEtran}
\bibliography{IEEEabrv, refs.bib}
\end{document}